\documentclass[%
reprint,
superscriptaddress,
groupedaddress,
nobibnotes,
secnumarabic,
amsmath,amssymb,
aps,
prb,
]{revtex4-1}

\usepackage{amsmath}
\usepackage{graphicx}
\usepackage{bm}
\usepackage[utf8]{inputenc} 
\usepackage{hyperref}
\usepackage{xcolor}

\newcommand{\angstrom}{\text{\normalfont\AA}}
\makeatletter
\newcommand*{\rom}[1]{\expandafter\@slowromancap\romannumeral #1@}
\makeatother
\begin{document}


\title{Collective excitations and plasmon spectrum in twisted bilayer graphene near the magic angle}

\author{Xueheng Kuang}
\affiliation{Key Laboratory of Artificial Micro- and Nano-structures of Ministry of Education and School of Physics and Technology, Wuhan University, Wuhan 430072, China}

\author{Zhen Zhan}
\email{zhen.zhan@whu.edu.cn}
\affiliation{Key Laboratory of Artificial Micro- and Nano-structures of Ministry of Education and School of Physics and Technology, Wuhan University, Wuhan 430072, China}

\author{Shengjun Yuan}
\email{s.yuan@whu.edu.cn}
\affiliation{Key Laboratory of Artificial Micro- and Nano-structures of Ministry of Education and School of Physics and Technology, Wuhan University, Wuhan 430072, China}

\date{\today}

\begin{abstract}
Twisted bilayer graphene with tiny rotation angles have drawn significant attention due to the observation of the unconventional superconducting and correlated insulating behaviors. In this paper, we employ a full tight-binding model to investigate collective excitations in twisted bilayer graphene near magic angle. The polarization function is obtained from the tight-binding propagation method without diagonalization of the Hamiltonian matrix. With the atomic relaxation considered in the simulation, damped and undamped interband plasmon modes are discovered near magic angle under both room temperature and superconductivity transition temperature. In particular, an undamped plasmon mode in narrow bands can be directly probed in magic angle twisted bilayer graphene at superconductivity transition temperature.  The undamped plasmon mode is tunable with angles and gradually fades away with both temperature and chemical potential. In practice, the flat bands in twisted bilayer graphene can be detected by exploring the collective plasmons from the measured energy loss function. 
\end{abstract}


\maketitle

\section{introduction}
Twisted bilayer graphene (TBG), one sheet of graphene rotates relatively to the other, has recently attracted extensive interests in the scientific community. In the TBG, one degree of freedom--the rotation angle--is introduced to tune its electronic properties. Experimental investigations are focusing on the properties of twisted bilayer graphene with rotation angle $1.05^\circ$--the so-called ``magic'' angle\cite{pnas2011moire}, where a plethora of fantastic phenomena, for instance, superconductivity\cite{cao2018unconventional,science2019tuning}, localized and correlated states\cite{cao2018correlated,nature2019fulltb,corre_insulatr2019electronic,supcorrelation2019superconductors,manybody2019spectroscopic}, charge-ordered states\cite{correlation2019chargeorder} and quantum anomalous Hall effect\cite{hall2020intrinsic}, are constantly observed. Recently, plasmons arising from interband collective excitations are detected by utilizing the scattering-type scanning near-field optical microscope(s-SNOM) at TBG with $1.35^\circ$\cite{plasexp2019cao}. Interestingly, plasmons occurring near magic angle could be used to mediate the unconventional superconductivity\cite{prs2020superconductivity}. 
Consequently, the importance of electron-electron interactions provokes us to gain insights into collective excitations in TBG, in particular near the magic angle. 

The full tight-binding (TB) model\cite{fulltb2010flat,full_tb2012numerical,full2017relax,full2018lammps,lammps2019continuum,arxiv2019relaxation} and simplified continuum model\cite{continu2007graphene,pnas2011moire,continu2012NJP,moon2013optical,continu2016fang,continu2018PRX,continu2019prb,continu2019prl,continu2019prs} are widely used to investigate the electronic properties of the TBG. Within the frame of these two models, the calculated properties of TBG can be in good agreements with experimental ones\cite{nc2013strain,strustrain2018prl,nature2019fulltb,zhan2020large,plasexp2019cao}. In general, if only low-energy properties are required, one could be prone to adopt simplified continuum models since the TBG near the magic angle contains over 10,000 atoms in its moir\'{e} unit cell. In this case, it is difficult to obtain eigenvalues and corresponding eigenstates via the diagonalization of the huge Hamiltonian matrix in the tight-binding model. For instance, optical conductivity and plasmonic properties of the TBG are studied by using the continuum model without considering the atomic relaxation\cite{moon2013optical,NJP2013optical,nano2016plas}. As indicated in both experimental and theoretical works, relaxation has non-negligible effects to the electronic properties of the TBG, especially for relatively small twist angles \cite{cao2016prl,continu2018PRX,continu2019prs,full2018lammps}. Therefore, after taking into account the out-of-plane relaxation, the continuum model is subsequently used to analyze plasmons in doped and undoped TBG \cite{pnasplas2019intrinsically,plasexp2019cao,arxiv2020plas}. 
Nevertheless, continuum models used in TBG are effective near small twist angles or at low energies. It is not a universal method to investigate a wide range of angles and to consider the local electronic environment near each atom. Moreover, the atomic relaxation implemented in continuum models are derived from the TB results{\cite{full2017relax,continu2018PRX,lammps2019continuum}. 
By using a tight-binding propagation method (TBPM) in the frame of a full tight-binding model, very large structures with the number of atoms up to hundreds of millions can be studied \cite{yuan2010tipsi}. The TBPM is based on the numerical solution of the time-dependent Schr\"{o}dinger equation with additional averaging over random superposition of basis states. More importantly, the relaxation effect, strain, physical defect, electric and magnetic field can be easily implemented in the model. Hence, as an atomic-scale approach, such a full tight-binding model without diagonalization is still meaningful to study electronic and optical properties of TBG at a large range of twist angles \cite{full2018ac}. 

\begin{figure*}[t]
	\includegraphics[width=0.9\textwidth]{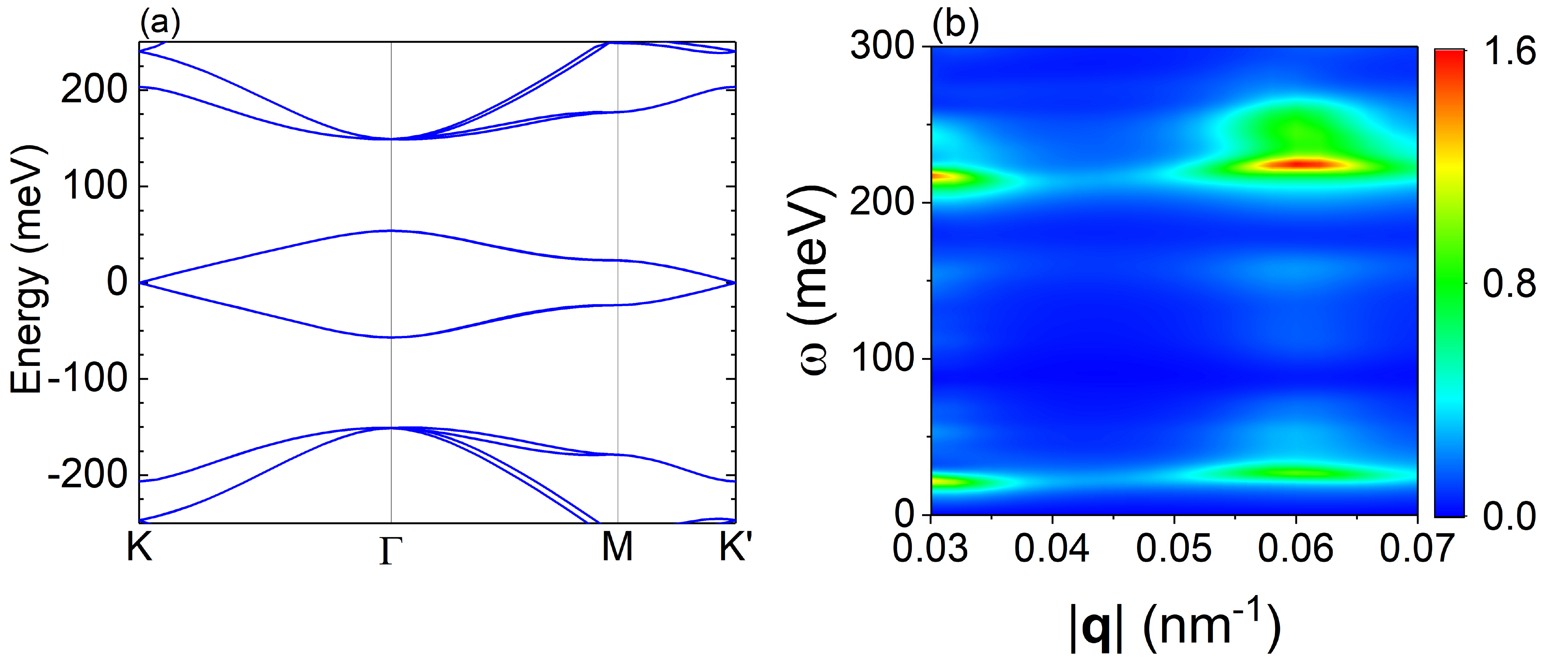}
	\caption{Reproduce the experimentally collective excitation for twisted bilayer graphene with twist angle $1.35^\circ$. (a) The band structure obtained from a tight-binding calculation. Similar to the result in Ref. \onlinecite{plasexp2019cao}, a band gap with the value of 100 meV is obtained by assuming that interlayer hoppings of A or B sublattices between two layers are reduced to zero. (b) The loss function $S=-\mathrm {Im}(1/\varepsilon)$ as functions of $\omega$ and $|\mathbf q|$. The plasmon mode above 200 meV is the one observed in the experiment\cite{plasexp2019cao}.}
	\label{fig:experi_loss}
\end{figure*}

In this work, beyond those aforementioned continuum models, we systematically investigate the plasmonic properties of TBG near magic angle by using the TBPM. 
We modulate the interlayer interaction of the relaxed TBG with twist angle $1.35^\circ$ to reproduce correctly the plasmons detected in a recent experiment\cite{plasexp2019cao}. Therefore, the Kubo formula implemented in the TBPM is accurate enough to investigate the relaxed TBG with tiny rotation angles. Then, plasmon spectra of TBG with rotation angles $1.35^\circ$ and $1.05^\circ$ for temperatures at both 300 K and 1 K are calculated. We observe that two collective plasmon modes are detected in twisted bilayer graphene near the magic angle, that is, a damped mode at the high energy range and an undamped one in the low energy range. The low-energy plasmons that are attributed to the collective excitations among the flat bands are sensitive to the temperature and doping effects. Such undamped plasmon is of keen interest for a bunch of applications, for instance, quantum information science and high-Q resonators.

\section{results and discussion}

In a recent experiment, an interband plasmon mode with the value around 200 meV was reported in the charge-neutral TBG near the magic angle\cite{plasexp2019cao}. The band structure of such system had a large band gap of 100 meV between the flat bands and the first excited band at the $\Gamma$ point of the Brillouin zone. Such unusual band gap could be explained by the extremely suppressed AA interlayer interaction due to the electron-electron interaction or extrinsic effects, for instance, the way samples are fabricated and the hBN encapsulation \cite{plasexp2019cao}. In fact, in our investigation of the magic angle in the Supplementary Information, the maximum band gap appears in the magic angle samples and has a value around 40 meV. In the Ref. \onlinecite{plasexp2019cao}, the suppression of AA interlayer interactions can be realized by reducing the inter-layer coupling in the AA regions $\mu_0$ in a continuum model. Here, we reproduce the large band gap by locally tuning the AA interlayer hopping in the full tight-binding model. In the superlattice cell with twist angle $\theta=1.35^\circ$, all the atoms can be separated into two sublattices, sublattice A and sublattice B. By changing the interlayer hopping parameter $t_1$ between A or B sublattice to 0 (such approximation resembles the transformation of effective AA and BB interlayer coupling $u_0$ to zero in the continuum model\cite{continu2018PRX}), we obtain a band gap of 100 meV as shown in Fig. \ref{fig:experi_loss}(a). The band structure shows great agreements with the one obtained from the continuum model in Ref. \onlinecite{plasexp2019cao}. The corresponding energy loss function Eq. (\ref{loss}) can be obtained by using the Kubo formula Eq. (\ref{kubo}) and (\ref{dielectric}) in our full tight-binding model. As shown in Fig. \ref{fig:experi_loss}(b), the loss functions are calculated within wavevectors $|\mathbf q|$ that are accessible and detected in experiment\cite{plasexp2019cao}. The plasmon mode takes place with energy near 210 meV, showing good agreements with the plasmon distribution in Ref. \onlinecite{plasexp2019cao}. Note that in the result in Fig. \ref{fig:experi_loss}, the lattice relaxations are taken into account in the simulation. We refer the readers to Supplementary Information for the atomic relaxation effects on the electronic properties of TBG. The extreme consistence with the experimental results verifies the accuracy and promising applications of our method.

\begin{figure*}[ht]
	\includegraphics[width=1\textwidth]{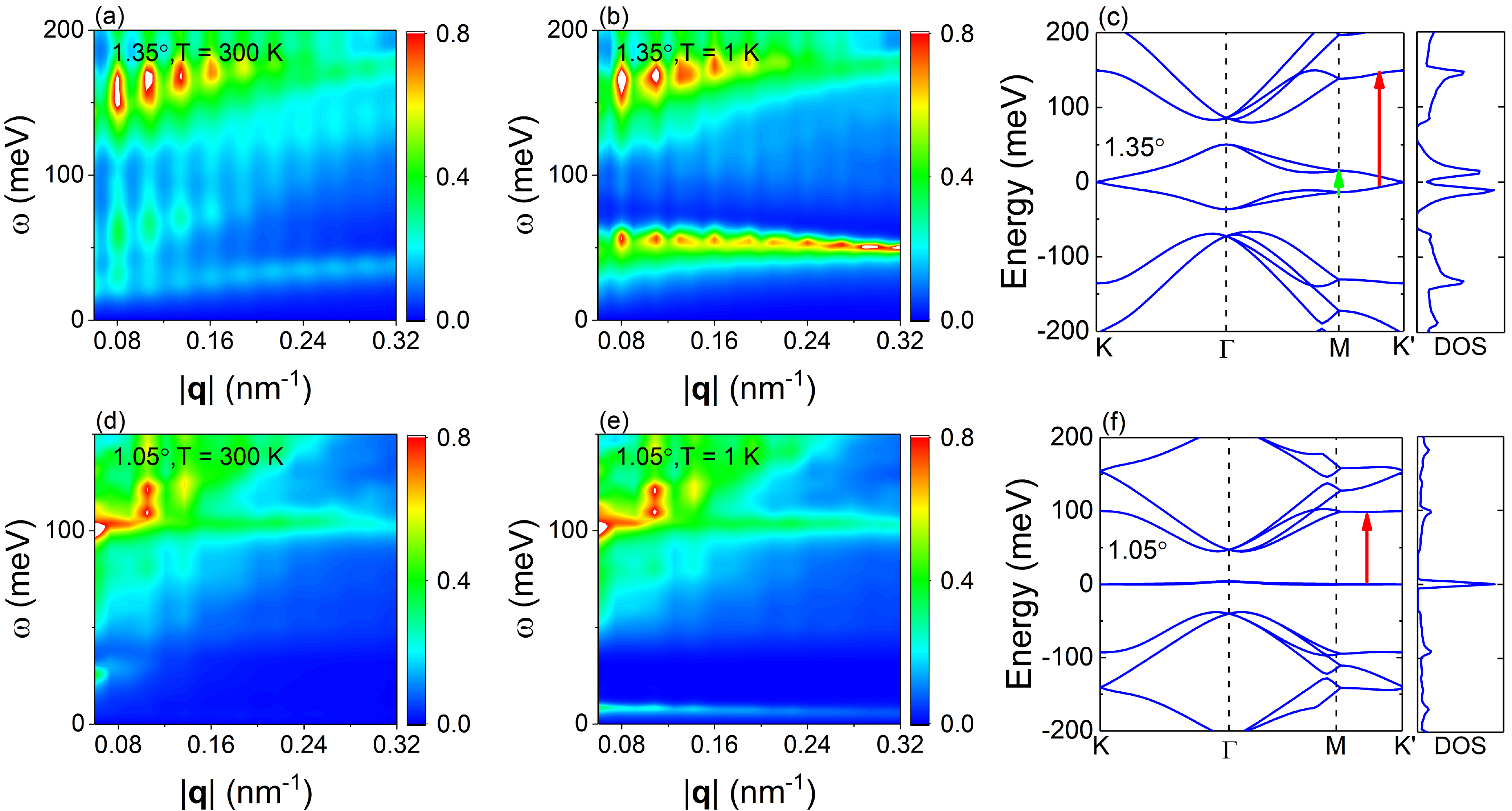}
	\caption{The loss function $-\mathrm{Im}(1/\varepsilon)$ change with the frequency $\omega$ and wave vector \textbf q for relaxed TBG with twist angle $1.35^\circ$ and $1.05^\circ$ at temperature (a),(d) $\mathrm T=300$ K and (b),(e) $\mathrm T=1$ K  , respectively. The possible interband transitions are indicated by red and green arrows in the band structure for (c) $\theta=1.35^\circ$ and (f) $\theta=1.05^\circ$. The wave vector is along the $\Gamma$-K direction. The chemical potential is $\mu=0$ and the background dielectric constant $\epsilon_B=3.03$ corresponds to the hexagonal boron nitride (hBN) substrate.}
	\label{fig: plas}
\end{figure*}

\begin{figure*}[htbp]
	\includegraphics[width=0.8\textwidth]{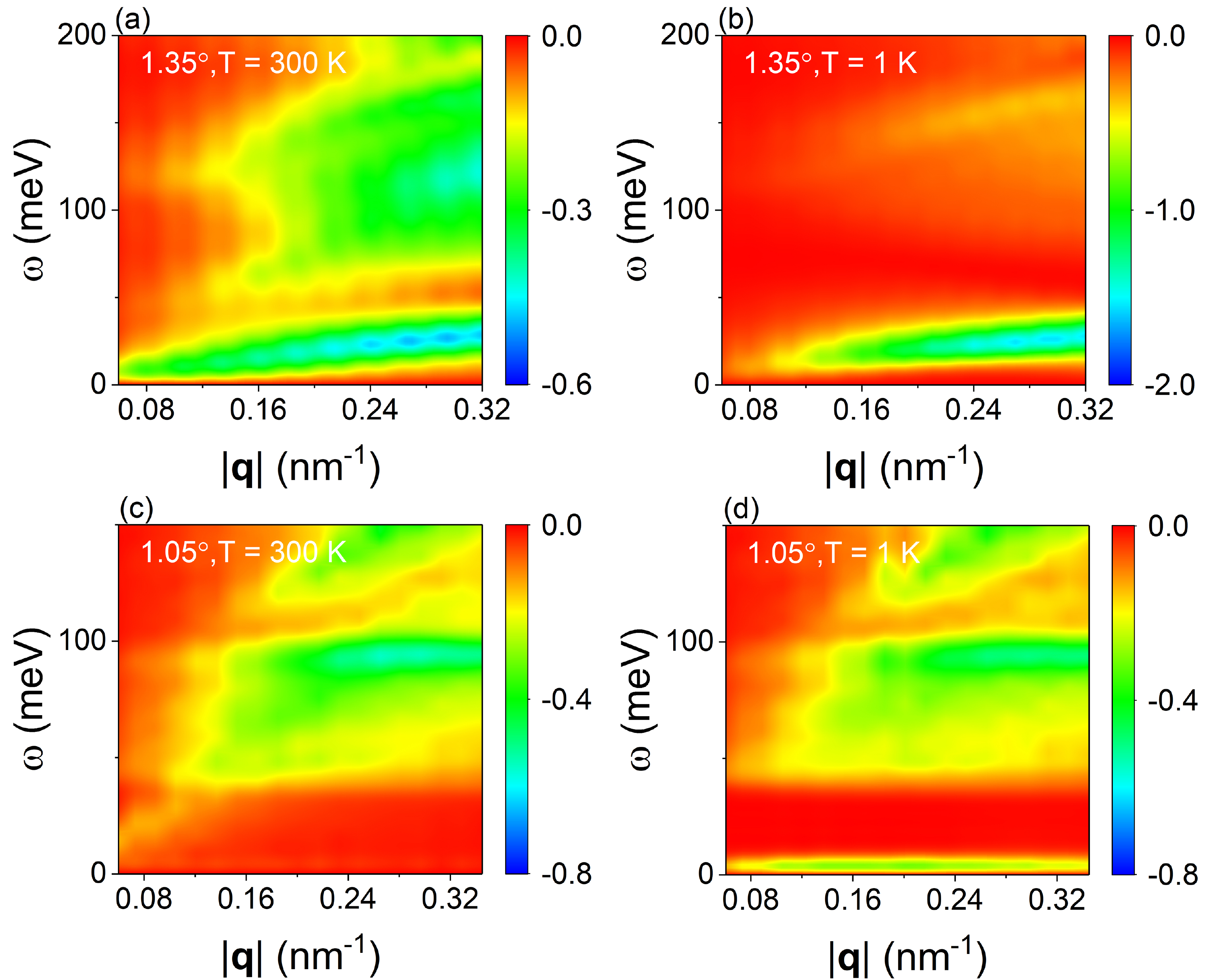}
	\caption{$-\mathrm{Im}\Pi(\textbf q, \omega)$ change with the frequency $\omega$ and wave vector \textbf q for relaxed TBG with $1.35^\circ$ and $1.05^\circ$ at (a),(c) T = 300 K  and (b),(d) T = 1 K, respectively. The wave vector is along the $\Gamma$-K direction. The chemical potential is $\mu=0$ and the background dielectric constant $\epsilon_B=3.03$ corresponds to the hBN substrate.}
	\label{fig:im_dyn}
\end{figure*}

Next, we utilize the Kubo formula in a full tight-binding model to theoretically explore collective excitations in TBG near the magic angle. Two samples with different rotation angles are created, one is with $\theta=1.35^\circ$ and the  other is the experimentally detected magic angle $\theta=1.05^\circ$. We explicitly discuss how to achieve the experimental magic angle $\theta=1.05^\circ$ with our full tight-binding model in the Supplementary Information. The atomic relaxations are taken into account in the calculations. Plasmon modes discovered from these excitations can be detected by experimental technologies, such as S-SNOM, electron energy loss spectroscopy (EELS). In experiments, when plasmon modes with frequency $\omega_p$ exist with low damping, which can be evaluated through Eq. \ref{damping},  the electron energy loss spectra possess sharp peaks at $\omega=\omega_p$. Here, the energy loss function $-\mathrm{Im}(1/\epsilon(\mathbf{q},\omega))$ is calculated with the dielectric function obtained from the Eq. (\ref{dielectric}) and the polarization function obtained from the Kubo formula in Eq. (\ref{kubo}). Plasmon spectrums of the TBG with two different rotation angles $1.35^\circ$ and $1.05^\circ$ are illustrated in Fig. \ref{fig: plas}. The corresponding band structures of the two angles are also plotted in Fig. \ref{fig: plas}(c) and (f), respectively.

For twisted bilayer graphene with $\theta=1.35^\circ$, interband plasmon modes close to 150 meV appear at both $\mathrm T=300$ K and $1$ K, which are attributed to the interband transitions from the valence band near the Fermi energy to the conduction band located at 150 meV (red arrow in Fig. \ref{fig: plas}(c)). These modes are similar to the ones around 200 meV in Fig. \ref{fig:experi_loss}. The 150 meV plasmon modes disperse within particle-hole continuum in Fig. \ref{fig:im_dyn}(a) and (b) with fast damping into electron-hole pairs. Other interband plasmon modes has the energy of 50 meV at temperature 1 K in Fig. \ref{fig: plas}(b), which can be interpreted as collective transitions between four bands near-zero energy (green arrow in Fig. \ref{fig: plas}(c)). As we can see from Fig. \ref{fig:im_dyn}(b), the $\textrm{Im}\Pi(\textbf q, \omega)$ is zero around 50 meV energy, showing that the 50 meV plasmons avoid Laudau damping from the interaction of collective excitations and single particle-hole transitions. 
At room temperature, the 50 meV plasmon modes split into two. One kind of modes has energy around 25 meV and the other above 50 meV which damps quickly with the increased wave vectors. As shown in Fig \ref{fig:im_dyn}(a), in contrast to the scenario at 1 K temperature, new single-particle transitions simultaneously occur around the plasmonic energy.  Obviously, at non-zero temperature, some states fluctuate around the Fermi energy, leading to the conduction band partially filled near the Fermi energy. Therefore, different from the result at 1 K, extra single-particle intraband transitions are induced at room temperture, giving rise to remarkable variations of the particle-hole continuum spectrum in Fig. \ref{fig:im_dyn}(a) and (b). 

For the TBG with $\theta=1.05^\circ$, the first obvious plasmon modes in Fig. \ref{fig: plas} locate at 100 meV at both T = 300 K and 1 K. These plasmons are also coming from the interband transitions illustrated in Fig. \ref{fig: plas}(f) (red arrow). The energy of the plasmons is smaller than the one discovered at $1.35^\circ$ since the van Hove singularity is located at the energy around 100 meV. These modes are damping ones as they cross the non-zero region in the particle-hole continuum spectrum in Figs. \ref{fig:im_dyn}(c) and (d), and it becomes clear with a fine and flat shape with momentum larger than $0.2\;\mathrm{nm^{-1}}$. At room temperature, different from the two splitting plasmons in TBG with $1.35^\circ$, the plasmons that stem from the collective transitions among four flat bands, vanish for TBG with $1.05^\circ$ in Fig. \ref{fig: plas}(d). Besides, single-particle transitions are almost not allowed in flat bands below 40 meV, corresponding to the value of band gap between the flat bands and the excited bands at $\Gamma$ point in Fig. \ref{fig: plas} (f), from which the continuum spectrum rises to non-zero zone in Fig. \ref{fig:im_dyn} (c). When the temperature declines to the critical temperature 1 K at which the superconductivity can be detected in the magic-angle system\cite{cao2018unconventional,science2019tuning}, a thin plasmon mode with energy 9 meV emerges and stretches to large q in Fig. \ref{fig: plas}(e), which is contributed to the collective excitations among flat bands ("flat-band plasmon"). Meanwhile, underneath the collective flat-band plasmon mode, the particle-hole transitions arise with occupying a tiny energy region ranging from 0 meV to 8 meV in Fig. \ref{fig:im_dyn}(d). As a result, this plasmon mode extends above the edge of this tiny energy zone and is free from the Laudau damping. 

The common feature of the interband plasmon modes in both cases is that they are ultra-flat over the whole $\textbf q$ vectors, which is in good agreement with the findings calculated by using the continuum model\cite{nano2016plas}. The higher energy plasmon mode can be changed at two different angles due to the variation of the Van Hove singularity with the twist angle, but rarely affected at different temperature. The low-energy plasmon mode originated from transitions between the four narrow-bands near the Fermi energy, was also observed in doped magic-angle TBG by using the continuum model\cite{pnasplas2019intrinsically}. Note that, in their work, the flat-band plasmon was detected by using a two-band toy model and the plasmon was independent of the temperature. However, in our tight-binding results, we only find the flat-band plasmons at low temperature.

Furthermore, It deserves to check if this flat-band plasmon is determined by the flatness of the four flat bands. As seen from the loss function in Fig. \ref{fig:mu_T}(a), for twisted bilayer graphene with three different rotation angles where flat bands appear (see Supplementary Information of Section \rom{3}), the position of the first peak, corresponding to the energy of the flat-band plasmon $E_p$, reaches the minimum energy for TBG with the magic angle $1.05^\circ$. The first peak of the loss functions for TBG with the other two angles $1.08^\circ$ and $1.02^\circ$ have the same higher energy since their bandwidths of flat bands are almost identical but both larger than that of the magic angle. What enlightens us here is that the different electronic response to various bandwidths could support us to unveil the flatness of flat bands in large moir\'e system where the calculations of the band structure becomes prohibitively expensive. 

                                                                                                                                                             \begin{figure*}[htbp]
	\includegraphics[width=1\textwidth]{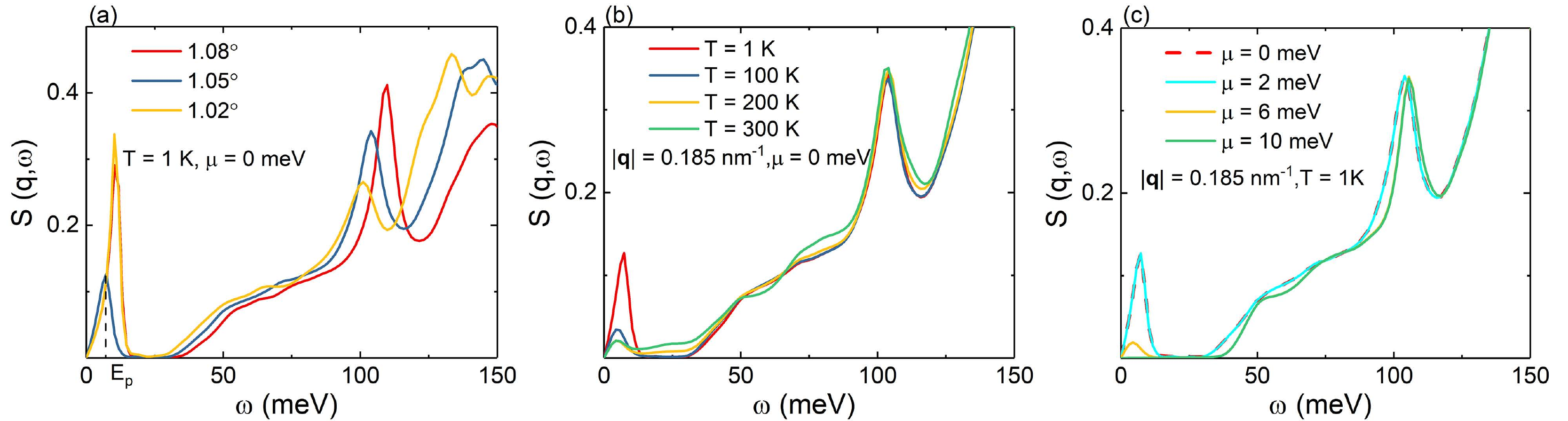}
	\caption{Loss function  $S(\mathbf{q},\omega)=-\mathrm{Im}(1/\epsilon(\mathbf{q},\omega))$ of twisted bilayer graphene (a) near $1.05^\circ$ at T = 1 K, (b) with $\theta=1.05^\circ$ at various temperatures and (c) with magic angle $1.05^\circ$ for different chemical potential $\mu$.}
	\label{fig:mu_T}
\end{figure*}                                                                                                                                                             

In the previous parts, effects of the temperature on plasmons have been studied briefly in undoped TBG. How will the flat-band plasmons of TBG with $1.05^\circ$ change with the temperatures varied in a large range from 1 K to 300 K pushes us to explore the loss functions under several temperatures, which is plotted in Fig. \ref{fig:mu_T} (b). The temperature suppresses significantly collective excitations between flat bands since the magnitude of the first peak declines sharply with increased temperatures. In the Ref.\onlinecite{pnasplas2019intrinsically}, dopped magic-angle TBG also has an undamped plasmon mode with energy around 7 meV, which is similar to aforementioned flat-band plasmons in our undoped case. Here, we further investigate the influence of the chemical potential on flat-band plasmons in twisted bilayer graphene with magic angle in Fig. \ref{fig:mu_T}(c). When the flat band is doped to a relatively low level of 2 meV and partially filled, the first peak of loss function is overlapped with the undoped one. The flat-band plasmon is not affected by the low-level doping. For a higher doping level, for instance, 6 meV that reaches the edge of flat bands, there is a weak electronic response with the first peak fading away. The filled flat bands at doping level 10 meV forbid all collective excitations from these flat bands. In addition, doping TBG to 6 meV has an equivalent effect to increasing temperature to 300 K in terms of the first peak of loss function. Therefore, it is essential to keep a low doping level or low temperature if one wish to observe the flat-band plasmon mode in TBG with magic angle.

In summary, collective excitations in twisted bilayer graphene are explicitly investigated by utilizing the Kubo formula in the frame of a full tight-binding model. We mainly compare plasmonic properties for TBG with angles $1.35^\circ$ and $1.05^\circ$ at different temperature. For the higher energy plasmon modes, they are conserved at both 300 K and 1 K and interact with single-particle electron-hole transitions, giving rise to a quick damping rate. Interestingly, undamped plasmon modes in the low energy range can be well defined since they go through the zero regions of the particle-hole spectrum. Increasing temperature from 1 K to 300 K turns the single undamped plasmon mode with lower energy into two damped bands at  $1.35^\circ$. At magic angle $1.05^\circ$, the undamped flat-band plasmon mode disappears at 300 K but emerges at 1 K with the minimum energy due to the narrowest bandwidth of flat bands. The angle dependence of flat-band plasmons near $1.05^\circ$ reflects different bandwidths of flat bands. The comparison of doping and temperature effects on the flat-band plasmonic peaks in loss function could guide us to study the flatness of flat bands in experiments. Lastly, in our simulation, the correct polarization functions are calculated with $\mathbf{q}$ inside Brillouin zone, whereas local field effect is supposed to be included when $\mathbf{q}$ approaches the zone boundary\cite{LFE1962quantum,LFE1963dielectric}. The effects are usually quite small but still a subject in the future. Recently, it has been theoretically predicted that unconventional superconductivity in TB mediated by the purely collective electronic modes\cite{prs2020superconductivity}. Therefore, a deep understanding of collective excitations in twisted bilayer graphene may also shed light on the ``plasmonic superconductivity''.

\section{methods}
A parameterized full tight-binding model with the commensurate approach to construct atomic structures of the twisted bilayer graphene, has been used to successfully reproduce the experimental findings \cite{strustrain2018prl,zhan2020large}. In this paper, we also use the commensurate method to construct the TBG with different rotation angle $\theta$\cite{zhan2020large}. In the tight-binding model, only $p_z$ orbitals are taken into account, we construct the Hamiltonian of the twisted bilayer graphene as
\begin{equation}\label{Hamil}
H = \sum_i\epsilon_i |i \rangle\langle i|+\sum_{\langle i,j\rangle}t_{ij} |i \rangle\langle j|,
\end{equation} 
where $|i\rangle$ is the $p_z$ orbital located at $\mathbf{r}_{i}$, and $\langle i,j\rangle$ is the sum over index $i$ and $j$ with $i \neq j$.  
According to the Slater-Koster (SK) formalism, the hopping integral $t_{ij}$, interaction between two $p_z$ orbitals located at $\mathbf{r}_{i}$ and $\mathbf{r}_{j}$ has the form\cite{SK1954simplified}
\begin{equation}
t_{ij}=n^2V_{pp\sigma}(r_{ij})+(1-n^2)V_{pp\pi}(r_{ij}),
\end{equation}
where $r_{ij}=|\mathbf{r}_{j}-\mathbf{r}_{i}|$ is the distance between $i$ and $j$ sites, with $n$ as the direction cosine along
the direction $\bm{e_z}$ perpendicular to the graphene layer . The Slater and Koster parameters $V_{pp\pi}$ and $V_{pp\sigma}$ follow
\begin{equation}
\begin{aligned}
V_{pp\pi}(r_{ij})=-t_0e^{q_\pi(1-r_{ij}/d)}F_c(r_{ij}),\\
V_{pp\sigma}(r_{ij})=t_1e^{q_\sigma(1-r_{ij}/h)}F_c(r_{ij}),
\end{aligned}
\end{equation}
where $d=1.42\; \angstrom$ and $h=3.349\; \angstrom$ are the nearest in-plane and out-of-plane carbon-carbon distance, respectively, $t_0$ and $t_1$ are commonly reparameterized to fit different experimental results\cite{lammps2019continuum,arxiv2019relaxation}. The parameters $q_\sigma$ and $q_\pi$ satisfy $\frac{q_\sigma}{h}=\frac{q_\pi}{d}=2.218 \angstrom^{-1}$, and the smooth function is $F_c(r)=(1+e^{(r-r_c)/l_c})^{-1}$, in which $l_c$ and $r_c$ are chosen as $0.265$ and $5.0\; \angstrom$, respectively. \\

\subsection{Density of states}
The density of states (DOS) for twisted bilayer graphene is calculated by using the TBPM\cite{yuan2010tipsi} in the frame of a full tight-binding model. The TBMP makes it possible to obtain the electronic properties for large-scale quantum systems, for instance, the DOS of TBG with rotation angle $\theta$ down to $0.48^\circ$\cite{zhan2020large} and of dodecagonal graphene quasicrystal\cite{yu2019dodecagonal} . The detailed formula of DOS can be expressed as
\begin{equation}\label{dos}
D(E)=\frac{1}{2\pi N}\displaystyle\sum_{p=1}^{N}\int_{-\infty}^{\infty}e^{iEt}\langle\varphi_p(0)|e^{-Ht}|\varphi_p(0)\rangle dt,
\end{equation}
where $|\varphi_p(0)\rangle$ is one initial state with the random superposition of basis states at all sites $N$. In all the calculations, the accuracy of the electrical properties can be guaranteed by utilizing a large enough system with more than 10 million atoms. For instance, in all the calculations, the number of the atoms in the supercell is set to 10 millions and the time steps is 4096.

\subsection{Dynamical polarization and dielectric function}
Dynamical polarization can be obtained by combining Kubo formula\cite{kubo1957statistical} with TBPM   as\cite{yuan2011kubo}
\begin{equation}\label{kubo}
\begin{aligned}
\Pi_K(\mathbf{q},\omega)=& -\frac{2}{S}\int_{0}^{\infty}dt e^{i\omega t}\mathrm{Im}\langle \varphi| n_F(H)e^{iHt}\\
&\times\rho(\mathbf{q})e^{-iHt}[1-n_F(H)]\rho(-\mathbf{q})|\varphi\rangle, 
\end{aligned}
\end{equation}
where $n_F(H)=\frac{1}{e^{\beta (H-\mu)}+1}$ is the Fermi-Dirac distribution operator, $\beta = \frac{1}{k_BT}$ being $T$ the temperature and $k_B$ the Boltzmann constant, and $\mu$ is the chemical potential; $\rho(\mathbf{q})=\sum_{i}c_i^{\dagger}c_i$exp$(i\mathbf{q}\cdot\mathbf{r}_i)$ is the density operator and $S$ is the area of the unit cell in TBG. The dynamical polarization function also can be obtained from the Lindhard function in a full tight-binding model as\cite{linear2005quantum}
\begin{equation}\label{Lindhard}
\begin{aligned}
\Pi(\textbf{q}, \omega) = &-\frac{g_s}{(2\pi)^2}\int_\mathrm{BZ}d^2\textbf{k}\sum_{l,l^{'}}\frac{n_\mathrm{F}(E_{\mathbf{k}l})-n_\mathrm{F}(E_{\mathbf{k'} l'})}
{E_{\mathbf{k}l}-E_{\mathbf{k'} l'}+\hbar\omega+\mathrm{i}\delta}\\
& \times |\langle \mathbf{k'} l'|\mathrm e^{\mathrm{i}\mathbf{q\cdot r}}|\mathbf{k}l \rangle |^2,
\end{aligned}
\end{equation}\\
where $|\mathbf{k}l \rangle$ and $E_{\mathbf{k}l}$ are the eigenstates and eigenvalues of the TB Hamiltonian Eq. (\ref{Hamil}), respectively, with $\mathit{l}$ being the band index, $\mathbf{k^{'}}$=$\mathbf{k}$+$\mathbf{q}$,  $\delta \rightarrow 0^+$, the integral is taken over the whole Brillouin zone (BZ), and the sum is calculated over all bands in the TB Hamiltonian Eq. (\ref{Hamil}). Note that, when the rotation angle goes down to $3.14^\circ$, the unit cell contains 1986 orbitals. The calculation of the polarization function from the Lindhard function becomes a numerical task due to the huge toll of diagonalization. On the contrary, in the Kubo formula in Eq. (\ref{kubo}), the exact diagonalization of the Hamiltonian is unnecessary. So we can compute the dynamical polarization for systems with tiny rotation angles\cite{zhan2020large}. The accuracy of the Kubo formula will pave the way to further investigate collective excitations of TBG near magic angle in a full tight-binding model. Therefore, in the following, to check the validity of the Kubo formula, we will compare the polarization function obtained from Kubo formula with the ones obtained from the Lindhard function. 

\begin{figure}[t]	
	\includegraphics[width=0.45\textwidth]{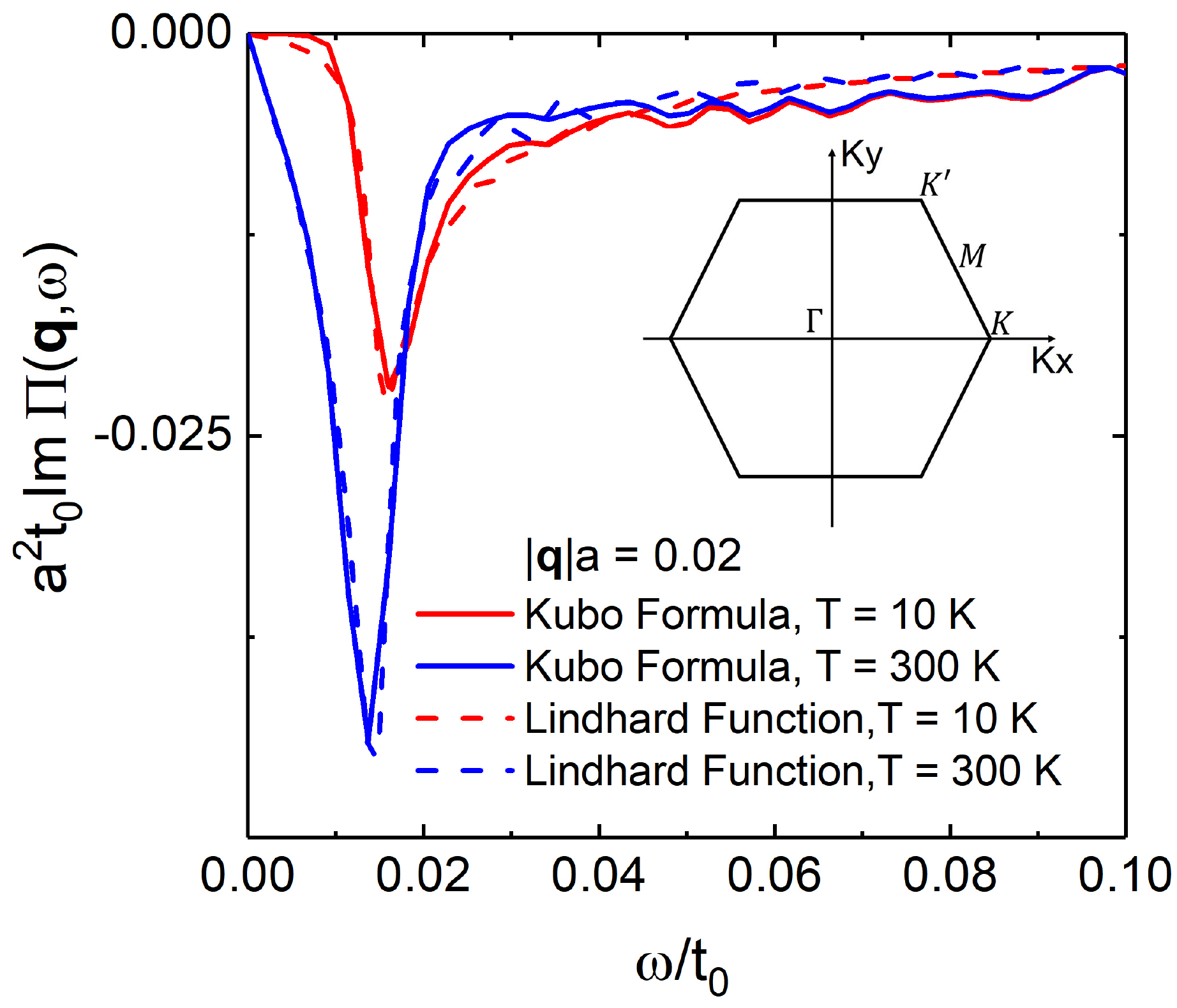}
	\caption{The imaginary part of the polarization function of rigid TBG with $\theta=6.01^\circ$ computed from Kubo formula (solid line) and from Lindhard function (dashed line) for $\mathrm T=10$ K (red line) and $\mathrm T=300$ K (blue line). The inset is the Brillouin zone of TBG. The wave vector $\mathbf{q}$ along $\Gamma$-K is $\mathrm |\mathbf q| a=0.02$ being $a=2.46\;\angstrom$ the lattice constant, and the chemical potential is $\mu=0$. The hopping parameters $t_0$ and $t_1$ are 2.78 eV and 0.33 eV, respectively.}
	\label{fig:6.01deg}
\end{figure}

With the polarization function acquired from either Kubo formula Eq. (\ref{kubo}) or Lindhard function Eq. (\ref{Lindhard}), the dielectric function that describes the electronic response to extrinsic electric perturbation, can be written within the RPA as 
\begin{equation}
\varepsilon(\textbf{q}, \omega) = 1- V(q)\Pi(\textbf{q}, \omega)
\label{dielectric}
\end{equation} 
in which $V(q)=2\pi e^2/{(\varepsilon_\mathrm{B}}{q})$ is the Fourier component of the two-dimensional Coulomb interaction, with $ \varepsilon_\mathrm{B}$ being the background dielectric constant. $\varepsilon_\mathrm{B}=1$ and $\varepsilon_\mathrm{B}=3.03$ represent the dielectric constant of the air and hexagonal boron nitride (hBN), respectively, in our calculations. The energy loss function can be expressed as
\begin{equation}
S(\mathbf q, \omega)=-\mathrm {Im}(1/\varepsilon(\mathbf q, \omega))
\label{loss}
\end{equation} 

To test the accuracy of the Kubo formula, we compare the polarization functions with that obtained from the Lindhard function at various angles and temperatures T. For twisted bilayer graphene with a large rotation angle, for instance, $\theta=21.78^\circ$, polarization functions obtained from the Lindhard function and the Kubo formula show remarkably quantitative agreement (not shown here). Furthermore, for smaller angles, the validity of the Kubo formula is tested in a smaller $\mathbf q$ that is located in the first Brillouin zone, see the inset in Fig. \ref{fig:6.01deg}. The results from the Kubo formula are still in good agreement with the ones from Lindhard function. Moreover, the temperature has an obvious impact on the polarization function, especially, for the imaginary part in the low-energy range which corresponds to the emergence of collective excitations at chemical potential $\mu=0$ eV. When the temperature is 300 K, excitations are induced in a tiny energy range smaller than $\omega/t_0=0.01$, from which excitations merely start at 10 K. The difference could be explained as a result of the modification of the electronic distribution near electric neutrality point at different temperature. These remind us that temperature can be a crucial factor to tune the collective excitations in TBG.

\begin{acknowledgments}
This work was supported by the National Science Foundation of China under Grant No. 11774269. Numerical calculations presented in this paper have been performed on the supercomputing system in the Supercomputing Center of Wuhan University.
\end{acknowledgments}

\bibliographystyle{apsrev4-1}
\bibliography{reference}

\end{document}